\documentclass[referee,useAMS,twocolumn,usegraphicx,usenatbib]{mn2e}
\usepackage{graphicx}
\usepackage{amssymb}
\usepackage{amsmath}
\usepackage{textcomp}
\usepackage{array}
\usepackage{multirow}
\usepackage{color}
\usepackage{times}

\newcommand{\ltsima}{$\; \buildrel < \over \sim \;$}

\title[The Main Sequence and the Fundamental Metallicity Relation]{The Main Sequence and the Fundamental Metallicity Relation in MaGICC Galaxies: Evolution and Scatter}
\author[A. Obreja et al.]{A. Obreja$^1$\thanks{E-mail: aura.obreja@uam.es}, C.~B. Brook$^1$, G. Stinson$^2$, R. Dom\'{\i}nguez-Tenreiro$^1$, \and  B.~K. Gibson$^{3,4}$, L. Silva$^5$ and G.~L. Granato$^5$\\
$^1$ Departamento de F\'{i}sica Te\'orica, Universidad Aut\'onoma de Madrid, E-28049 Cantoblanco Madrid, Spain\\
$^2$ Max-Planck-Institut f\"ur Astronomie, K\"onigstuhl 17, 69117, Heidelberg, Germany\\
$^3$ Institute for Computational Astrophysics, Dept of Astronomy \& Physics, Saint Mary's University, Halifax, NS, B3H 3C3, Canada\\
$^4$ Jeremiah Horrocks Institute, University of Central Lancashire, Preston, PR1 2HE, United Kingdom\\
$^5$ Osservatorio Astronomico di Trieste, INAF, Via Tiepolo 11, I-34131 Trieste, Italy\\
}

\begin{document}

\date{\today}

\pagerange{\pageref{firstpage}--\pageref{lastpage}} \pubyear{}

\maketitle

\label{firstpage}

\begin{abstract}
Using cosmological galaxy simulations from the MaGICC project, we
study the evolution of the stellar masses, star formation rates and gas phase
abundances of star forming galaxies. We derive the stellar masses and
star formation rates using observational relations based on spectral energy
distributions by applying the new radiative transfer code GRASIL-3D to our
simulated galaxies. The simulations match well the evolution of the stellar
mass-halo mass relation, have a star forming main sequence that maintains a
constant slope out to redshift z $\sim$ 2, and populate projections of the stellar
mass - star formation - metallicity plane, similar to observed star forming disc
galaxies. We discuss small differences between these projections in observational 
data and in simulations, and the possible causes for the discrepancies.
The light-weighted stellar masses are in good agreement with the simulation values, 
the differences between the two varying between 0.06 dex and 0.20 dex.
We also find a good agreement between the star formation rate tracer and
the true (time-averaged) simulation star formation rates.
Regardless if we use mass- or light-weighted quantities, our simulations 
indicate that bursty star formation cycles can account for the
scatter in the star forming main sequence.

\end{abstract}


\section{Introduction}
\label{intro}

The manner in which galaxies evolve and grow over their lifetime is reflected in the observed correlation between their star formation rate (SFR) and their assembled stellar mass, 
dubbed the \textit{main sequence} (MS) \textit{of star formation}. 
In the local Universe, the SFRs and stellar masses of star forming galaxies correlate tightly \cite[e.g.]{Brinchmann:2004,Elbaz:2007,Noeske:2007}, and the relation has also been shown to hold at higher redshifts  \cite[][among others]{Daddi:2007, Karim:2011, Bouwens:2011, Wuyts:2011, Whitaker:2012}. 
The slope of the main sequence remains relatively constant with redshift, with  the zero point evolving 
in the sense that high-redshift galaxies form stars at a higher rate than local galaxies of similar stellar mass \citep{Wuyts:2011, Whitaker:2012}. 
The scatter around the main sequence shows little dependence  on mass, nor on redshift. There is a hint of  increased scatter with decreasing stellar mass \citep[e.g.][]{Whitaker:2012}.

Complementary to the locus of star forming galaxies on the main sequence, 
the relation between the mass and metallicity of galaxies provides further insights and constraints on the history and evolution of galaxies.  
The mass-metallicity (M-Z) relation connects stellar masses to both the gas phase \citep[e.g.][]{Garnett:2002, Tremonti:2004}    
and stellar metallicities \citep{Cowie:2008, Perez-Montero:2009}, 
and it has been observed over a range of redshifts \citep{Tremonti:2004, Erb:2006, Kewley:2008, Maiolino:2008, Zahid:2011, Henry:2013a, Henry:2013b, LaraLopez:2013}. 
At all $z$s the lower mass systems have lower metallicities. 
We note here that care must be taken in interpreting and reconciling these results, 
as the evolution of the relation is subject to differences in the observational techniques used at different redshifts and to calibrations of their abundances based on observed 
rest-frame optical line ratios \citep[e.g.][]{Zahid:2011}.

The star forming main sequence and mass metallicity relation have been combined in \textit{the fundamental metallicity relation} (FMR), relating the stellar mass, 
gas-phase metallicity and the SFR of galaxies \citep{Mannucci:2010, LaraLopez:2010}. 
The FMR reflects the cycle of inflows that feed star formation, the production of metals and enrichment of the interstellar medium (ISM), 
and the subsequent outflows that are driven by the energy released by the forming stellar populations \citep[see][]{Dave:2012,Dayal:2013,Lilly:2013}. 

These indicators of galaxy formation and growth provide important constraints on galaxy formation models. 
Perhaps the most ambitious of these models are those that include baryonic physics, 
employing hydrodynamics to follow the evolution of gas and star formation within a fully cosmological context, and allowing a detailed description of the structural, 
chemical and dynamical properties within the galaxies \citep{Katz:1992,Steinmetz:1994,Katz:1996}. 

Certainly, significant progress has been made with such models in recent years. 
In particular, simulations that regulate star formation in order to be more consistent with the empirical relation between stellar mass and halo mass \citep{Moster:2010, Guo:2010} 
are having success in forming more realistic galaxies \citep{Brook:2011, Guedes:2011, Brook:2012,  McCarthy:2012, Munshi:2013, Aumer:2013, Marinacci:2013}. 
A range of different feedback schemes have been used in these studies, and in particular many augment feedback from supernovae 
with feedback from massive stars prior to their explosion as supernovae \citep[][]{Stinson:2013}. 
Suffice to say that the recent simulations all input significantly more energy than was generally used in previous generations, 
which generally suffered from severe loss of angular momentum \citep[see, however,][]{Saiz:2001,Domenech:2012}, 
and too much mass in the  central regions \cite[e.g.][]{Navarro:2000,Piontek:2011}.

High resolution MaGICC `zoom' simulations have been shown to match the stellar-halo mass relation over
a wide range in mass, as well as a variety of scaling relations at z = 0, including those between luminosity, 
rotation velocity, size, HI gas content and metallicity \citep{Brook:2012,Stinson:2013}. 
No scaling with mass for the outflows, no mass-loading nor direction of outflows
is input by hand in our simulations, so any scalings with galaxy mass arises naturally from the energy feedback
implementation. These simulations also have slowly rising rotation curves and appropriately large disc-to-total ratios.

For redshifts z $>$ 2, using relatively low resolution simulations where hydrodynamics was used throughout a volume of 114 Mpc$^3$,
the simulated galaxies have also been shown to match the observed slope and scatter in the stellar mass - halo mass relation 
and galaxy stellar mass function for galaxies with stellar mass $M_{*} > 5\times10^{10} M_{\odot}$ \citep{Kannan:2013}.

We now further test our suite of high resolution `zoom' MaGICC simulated galaxies by studying their evolution in the parameter space of halo mass, stellar mass, 
star formation rate and gas phase metal abundance, and we discuss our results in comparison with observational data. 
In what can be considered an accompanying paper \citep{Brook:2013}, 
we have characterized the baryon cycle by quantifying the inflows, outflows and recycling rates of gas and metals into, and out of, 
the virial radius and star forming regions of our simulations. The evolution of the properties of stellar mass, 
star formation rates and metallicity that we follow in this current paper are a direct reflection of this baryon cycle.

The paper is outlined as follows:
in Section~\ref{sims} we recap the essential features of our simulations. In Section~\ref{pp-grasil} we outline how we post-process our simulations using 
GRASIL-3D \citep{Dominguez:2014} and derive stellar masses and star formation rates from synthetic spectral energy distributions 
that include the effects of dust. 
The evolution of the stellar mass-halo mass relation is shown in Section~\ref{abundance_match_evol}, 
followed by the evolution of the star forming main sequence and its scatter in Sections~\ref{main-sequence-evol} and~\ref{scatter}, respectively.
The mass-metallicity and the fundamental metallicity relation are presented in Sections~\ref{msfrmet} and~\ref{massmet}. 
Similarities and differences with observational data are highlighted in each of these sections. Section~\ref{discuss} summarizes our conclusions.

\begin{figure*}
\centering
\includegraphics[]{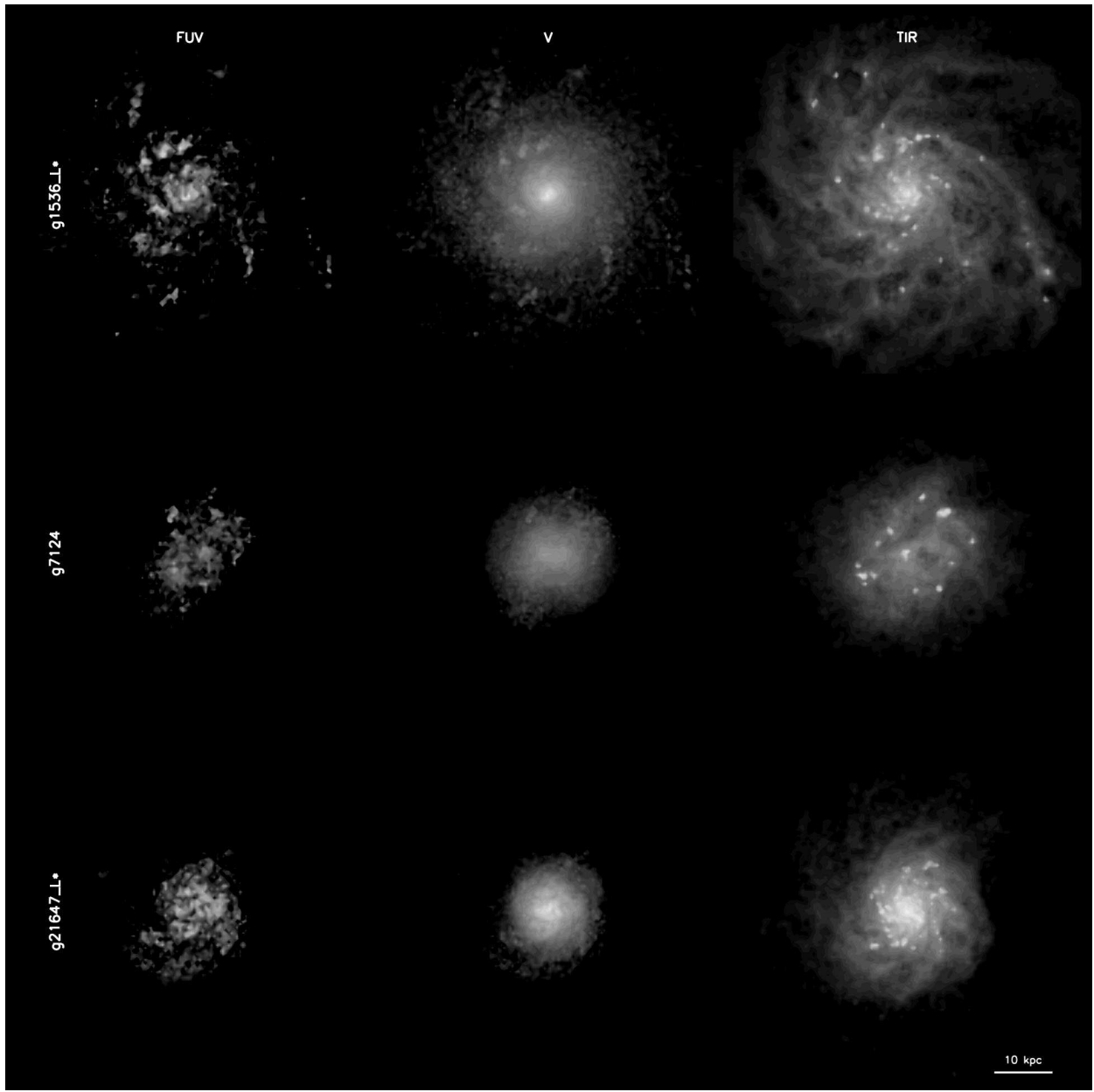}
\caption{
Example of GRASIL-3D images in the spectral ranges of far-UV, visible and total-IR (from left to right) for three of the more massive simulated galaxies at $z = 0$.}
\label{image}
\end{figure*}

\begin{table}
\begin{minipage}{3.0in}
\centering \caption{Simulation data where $m_{gas}$ is the initial mass of gas particles, h$_{soft}$ the minimum SPH smoothing length,
 and $M_{vir}$ and $M_{star}$ the total and stellar masses within the virial radius \citep{Bryan:1998} at $z= 0$.}
\begin{tabular}{l cccc}
\hline
Name & $m_{gas}$ & $h_{soft}$ & $M_{vir}$ & $M_{star}$\\
     & ($10^{4}$M$_{\odot}$) & (pc) & ($10^{10}$M$_{\odot}$) & ($10^{9}$M$_{\odot}$)\\
\hline
g21647\_L$^{*}$ & 18.9 &312 & 79.18 & 25.11 \\
g1536\_L$^{*}$  & 18.9  & 312  &  67.95 &  23.61 \\
g7124 & 18.9 &312 & 41.80 & 6.29 \\
g15784 & 7.5 &156 & 16.42 & 4.26 \\
g15807 & 7.5 &156 & 27.81 & 1.47 \\
g1536-Irr & 7.5 &156 & 7.71 & 0.45 \\
g5664 & 7.5 &156 & 5.92 & 0.24 \\
g21647-Irr & 7.5 &156 & 9.18 & 0.20 \\
\hline
\end{tabular}
\label{table1}
\end{minipage}
\end{table}

\section{The simulations}
\label{sims}

We analyze eight cosmological zoom simulations from the MaCICC project \citep{Brook:2012, Stinson:2013}, run with GASOLINE \citep{Wadsley:2004}, a parallel SPH tree-code. 
Initial conditions are derived from the McMaster Unbiased Galaxy Simulations \citep[][]{Stinson:2010}. 
The simulations include metal line cooling \citep{Shen:2010}, the effect of a ultraviolet ionizing background, supernova and early stellar feedback. 
We describe here the most important implementations and refer the reader for details of the code to \cite{Wadsley:2004}, and of the feedback implementation to \citep[][]{Stinson:2010}.

Stars are produced with a star formation rate $\propto \rho^{1.5}$ from cool ($T$$<$$15000$\,K) and dense ($n_{th}$$>$9.3\,cm$^{-3}$) gas, 
with a star formation efficiency parameter of 0.017.
The blastwave formalism \citep{Stinson:2006} is used to implement supernova feedback, at the end of massive stars (8\,M$_\odot$) lifetime, 
$10^{51}$ erg of energy being deposited into the surrounding medium. 
Prior to their explosion as SNe, massive stars inject energy into the surrounding gas. 
This feedback mechanism has also been included \citep{Stinson:2013} as pure thermal energy feedback in order mimic the weak coupling to the surrounding gas \citep{Freyer:2006}.
Thermal energy feedback is highly inefficient in these types of simulations \citep{Katz:1992,Kay:2002}, $\sim$90\% of the injected energy being rapidly radiated away. 
Thermal energy feedback is highly inefficient in these types of simulations \citep{Katz:1992,Kay:2002} due to our inability to resolve the ISM of star forming regions, 
with a significant amount of  the injected energy radiated away prior to the next simulation timestep, 
resulting in an  effective coupling of early stellar feedback to the ISM.

A Chabrier IMF \citep{Chabrier:2003} is used to compute the ejected mass and metals, which are distributed among the nearest neighbor
gas particles according to the smoothing kernel \citep{Stinson:2006}. Metal abundances are based on the SNII yields of \cite{Woosley:1995} and SNIa yields of \cite{Nomoto:1997}.
The turbulent metal mixing is mimicked using a shear-dependent diffusion term \citep{Shen:2010}. Cooling rates are computed by taking into account the diffused metals.

In Table~\ref{table1} we give the stellar and virial masses at $z = 0$, the SPH smoothing lengths and the initial mass of gas particles for 
the simulated galaxies.

\begin{figure*}
\centering
\includegraphics[]{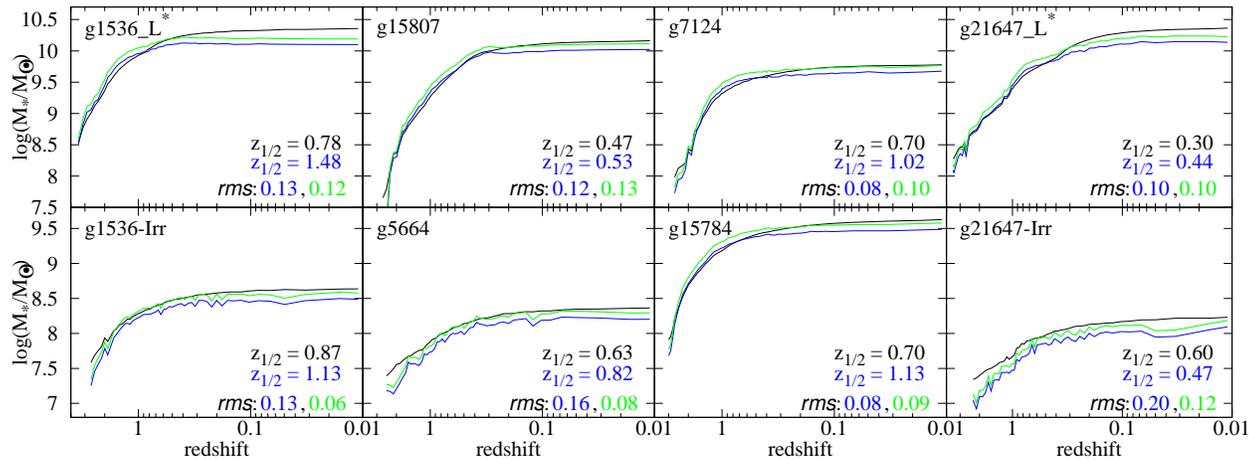}
\caption{
Stellar mass assembly for the simulated galaxies as predicted by simulations (black curves), and as derived using the B-V colors and mass-to-light ratios in the B-band (blue curves) and 
V-band (green curves) respectively. The half mass redshift, $z_{1/2}$ is given in each case for the black and blue curves. The \textit{rms} scatter of the stellar mass 
derived using the B-V colors and the mass-to-light ratios in the B- and V-bands with respect to the \textit{real} mass are also shown.}
\label{mass_evol}
\end{figure*}

\begin{figure*}
\centering
\includegraphics[]{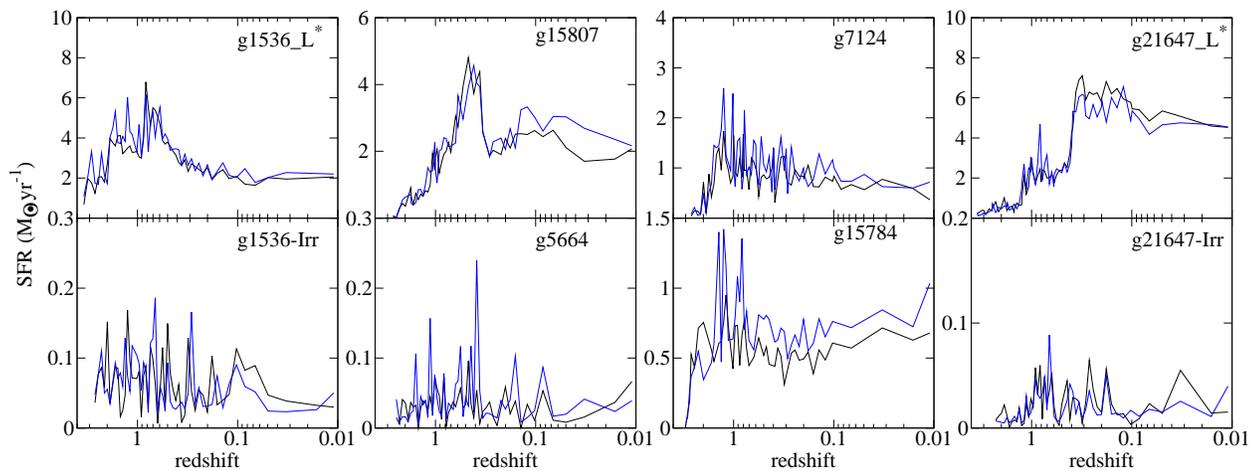}
\caption{
SFR histories for the simulated galaxies as predicted by simulations (black curves), and as derived using the infrared-corrected FUV luminosities (blue curves).
The \textit{observational} SFRs are fairly consistent with the \textit{real} ones.}
\label{sfrh_evol}
\end{figure*}

\section{Post-processing with GRASIL-3D}
\label{pp-grasil}

In order to make a more meaningful comparison with observational data, we model the light-weighted quantities for stellar masses and star formation rates. 
For this purpose, we used the radiative transfer code, GRASIL-3D \citep{Dominguez:2014}, which is capable of generating Spectral Energy Distributions (SEDs) for 
simulated galaxies of arbitrary geometries, taking into account the dust processed light from stellar populations. These calculations assumed the same Chabrier IMF as used 
in the simulations, and the stellar library from \cite{Bruzual:2003}. We use a  parameter set that governs the destruction 
of molecular clouds by the embedded stars which leads to fluxes from visible to IR in the ranges spanned by observational data samples.
This parameter set has escape time-scale from molecular clouds of $t_{0} = 10 Myrs$, radius for the clouds of $r_{mc} = 14 pc$, 
a cloud threshold density of $\rho_{mc,thres} = 3.3\times10^{9} M_{\odot}kpc^{-3}$ and a log-normal probability distribution function \citep{Wada:2007} with $\sigma = 3.0$, 
while the mass of a single cloud is assumed to be $m_{mc} = 10^{6} M_{\odot}$.

We construct GRASIL-3D inputs for all the snapshots with $z < 3.5$ of the eight simulated galaxies, and obtained the face-on 
surface brightness profiles in the r-band. We then  derive the Petrosian radii, $R_{P}$ \citep{Blanton:2001}, which we use to 
define the radial limits of our simulated galaxies (\textit{R}$_{lim}$=2$R_{P}$).
GRASIL-3D calculations have been done in the rest frame of the objects. 
Luminosities, fluxes and colors have been computed from the integrated face-on SEDs  within R$_{lim}$.
Therefore, our sample consists in eight galaxies \textquoteleft observed\textquoteright at various times in their evolution. 
In Figure~\ref{image} we show an example of GRASIL-3D images, in relevant bands for this study for three of the more massive simulated galaxies at $z = 0$.
  
All post-processing with GRASIL-3D has been done using fixed parameters, 
which are nevertheless suitable for both starburst and quiescent galaxies \citep{Dominguez:2014}.
Also, all observables have been computed in the rest-frame of the objects, and no detector properties apart from the transmissions in each band have been taken into account. 
Since we use broadband fluxes from far-UV to far-IR to estimate stellar masses and SFR, over a wide range of mass, these limitations are important. 
Technically, at high redshift, our smallest galaxies could not be observed in all bands we employ.
Therefore, our comparison of light and mass weighted quantities is an idealized one.
Nevertheless, the process of viewing the simulations in a manner analogous to observers helps to build confidence in the comparisons we make between observations and simulations.

There are various ways in which intrinsic properties like mass and SFR are inferred from observables. In the case of stellar masses, the current most widely used method is SED fitting, 
which became feasible with the recent advance in accurate multi-wavelength data and the refinement of stellar population synthesis models to include effects like 
post-main sequence stellar evolution. 
We decided to use simple color dependent mass-to-light relations, which nevertheless have been tested against other methods, like the Tully-Fisher one. 
The recent work of \cite{McGaugh:2013} provides color dependent mass-to-light relations in four different bands (B, V, Ks and IRAC 3.6$\mu$m). 
We use  the B- and V-bands mass-to-light ratios, which return masses closer to the simulations values than the ones in the Ks and IRAC 3.6$\mu$m bands.

Global SFRs can be inferred using emission lines or broadband fluxes \cite[see][for a review on SFR tracers]{Kennicutt:2012}.
In our study we employ the far-UV tracer of \cite{Hao:2011}, where the FUV flux is corrected for dust obscuration using the total IR emission (TIR). 
These tracers reflect the SFR over timescales of $\sim$100 Myrs, which are well resolved by our simulations. 
Therefore, mass-weighted SFR are time-averaged over 100 Myrs. 
One thing to keep in mind is that a good estimation of the TIR flux is observationally difficult, especially for dwarf galaxies.

We firstly checked whether our galaxies fall within the $B-V$ and $M_V$ ranges spanned by the data sample used by \cite{McGaugh:2013}. 
We also checked the ranges spanned by our objects in the $IRX=log(L_{TIR}/L_{FUV,obs})$ vs. $FUV-NUV$ diagram (Figure 3 of \cite{Hao:2011}),
finding that the majority of our objects have $IRX$ within $[−0.19,2.39]$ and $FUV-NUV$ within $[0.07,1.06]$, and moreover fall close
the curve defined by the observational sample employed by \cite{Hao:2011}, albeit with a large dispersion (0.46~dex).
While the observational sample contains only local normal star forming galaxies, 
some of the simulations snapshots correspond to peaks of star formation and some of the objects are dwarfs, 
cases outside of the sample used for the calibration.
The error in the IRX-based attenuation correction for starburst is, in any case, $< 0.3$~mag \citep{Hao:2011}. On the other hand, this prescription when applied to dwarfs might suffer from problems.
Indeed, the highest rate of discrepancy in the  $IRX$ vs. $FUV-NUV$ occurs for the two lowest mass simulations (g5664 and g21647-Irr), 
which have a significant number of snapshots with $IRX < −0.19$ (35\% and 56\%, respectively). However, even in these two cases, the synthetic SFR closely follows the real one.

We shall be calling the whole sample \textit{complete}, and the corresponding restriction to the observational
ranges of \cite{Hao:2011} and \cite{McGaugh:2013}, \textit{limited}. 
There is no significant difference in our results between these two samples, as can be seen in consistency between the light- and mass-weighted stellar mass and SFR in Figures~\ref{mass_evol} and \ref{sfrh_evol}.

In Figure~\ref{mass_evol} we give the stellar mass assembly tracks for the simulated galaxies. 
The black lines are obtained by summing the stellar mass within $R_{lim}$; 
the colored lines are obtained by applying the mass tracers from \cite{McGaugh:2013}, 
with blue (green) based on the M/L ratio in the B-band (V-band). 
The two light-weighted stellar masses follow quite closely the true simulation stellar mass, the difference being no larger than 0.20 dex. 
Figure~\ref{mass_evol} indicates that the most massive galaxies at low redshifts are better traced by the V-band derived masses, while at higher z the B-based stellar mass gives slightly better results.
On the other hand, for the lower mass systems (g1536-Irr, g5664 and g21647-Irr), the V-band is a better tracer than the B-band at all zs.

The differences between the light-weighted (blue) and mass-weighted (black) SFRs is shown in Figure~\ref{sfrh_evol}, 
where light-weighted SFRs are derived from IR-corrected FUV luminosities \citep{Hao:2011}. 
This SFR tracer is based on the correlation between the IR/UV luminosity ratio and UV color, which is used to estimate the dust attenuation. 
The mass-weighted SFRs were computed by counting all the 
stellar particles that formed in the last 100 Myrs of each simulation snapshot. This time interval for averaging has been used in order to be 
consistent with the SFR tracer. The observational  tracer follows closely the bursty simulation  SFR.
The differences between the \textquoteleft observed\textquoteright and \textquoteleft true\textquoteright ~SFRs are linked to the assumptions used
when constructing the tracer. For example, the attenuation-corrected FUV luminosities of \cite{Hao:2011} are consistent with 
evolutionary synthesis models which assume a constant solar metallicity. 

The mean fraction of SFR coming out in IR (averaging over all the snapshots of a simulated galaxy) increases linearly with log(M$_{*}$) at $z = 0$.
At the low mass end ($M_{\star}$$\sim$10$^8$M$_{\odot}$), the IR flux contributes 25\% to the global SFR, increasing to 80\% of the total SFR for $M_{\star}$$\sim$10$^{10}$M$_{\odot}$.
For the three most massive galaxies (g1536\_L*, g15807 and g21647\_L*), this fraction increases quite smoothly with decreasing $z$, while
for the less massive galaxies it is highly variable.

Computing light-weighted quantities from face-on images means we are minimizing the effect of dust obscuration. 
Therefore, we checked the ranges of our observables when different inclinations are considered (face-on to edge-on). 
For g7124, g1536-Irr, g5664, g15784 and g21647-Irr, the derived mass from edge-on images is \ltsima15\% less than the face-on values. For the massive 
galaxies g1536\_L* and g15807, the mass decrease can be as high as 25\%.   g21647\_L* has the most extreme inclination effect, with differences between edge-on and face-on up to 50\%,  
although this only occurs at low redshift $z < 0.4$. 
For larger redshifts, the effect of inclination for the three massive galaxies is typically  $\sim$20\%.

The SFR tracer is less affected by inclination than the mass tracers: typical values of the decrease in derived SFR due to inclination   are $\sim$10\%. 
The effects are slightly lower at  higher  redshift, and show no clear trend with mass.

\begin{figure}
\centering
\includegraphics[]{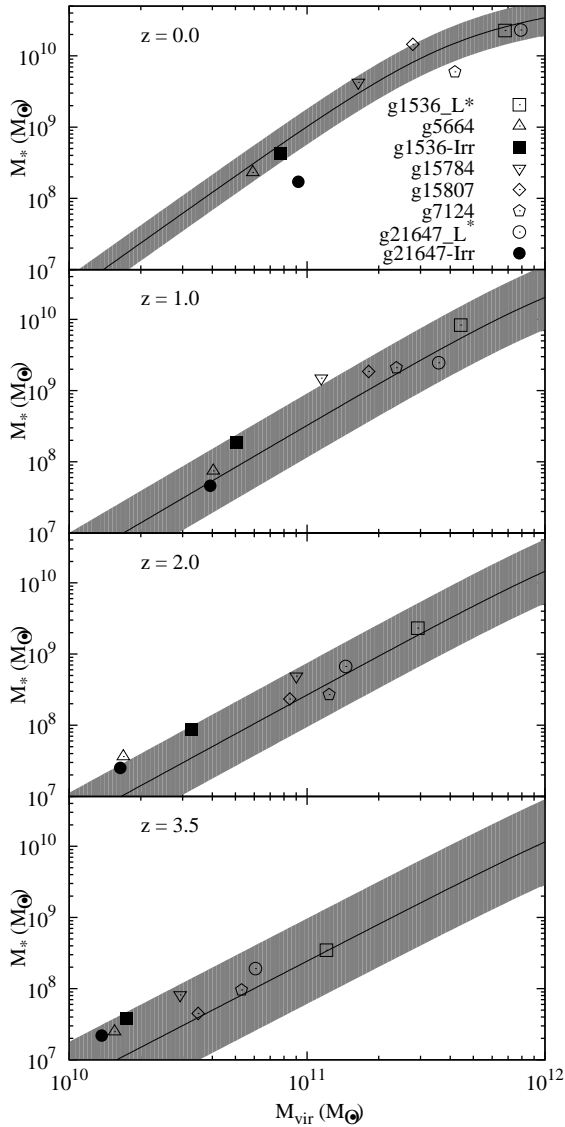}
\caption{
Stellar mass versus halo mass for the simulated galaxies, at four redshifts. The black curves are the abundance matching functions derived by \citealt{Moster:2013},
with their corresponding 1$\sigma$ deviations (grey areas).}
\label{abundancematching}
\end{figure}

\section{Results}

The effect of different star formation and feedback parameters and implementations applied to a single Milky Way like galaxy, g1536\_L*, has been studied by \cite{Stinson:2013}. 
The variation found in the evolution of star formation pointed to the need for feedback from massive stars prior to their explosion as supernovae.  
Using the same feedback scheme, a large sample of high mass simulated galaxies was also shown  to match the relation at high redshift \citep{Kannan:2013}. 
Here we test the effectiveness of the implemented feedback by exploring the evolution of our suite of high resolution zoom simulations that span a wide range in stellar mass.

\subsection{Evolution of the stellar-halo mass relation}
\label{abundance_match_evol}

In Figure~\ref{abundancematching}, we show the stellar mass - halo mass relation of our simulated galaxies over plotted on the empirical results from \cite{Moster:2013}, 
along with their 1$\sigma$ scatter,  at four redshifts. 
The galaxies follow the same evolutionary path of stellar mass assembly  as those inferred by the abundance matching techniques used in  \cite{Moster:2013}.  
The stellar masses in this figure are those computed directly from the simulations, but as seen above there is little difference if light-weighted stellar masses are used. 

Our suite of simulated  galaxies that include early stellar feedback from massive stars, as well as feedback from supernovae,  not only resemble observations in the Local Universe, 
but also have stellar masses that evolve with redshift in a manner that resembles real galaxies.

\begin{figure*}
\centering
\includegraphics[]{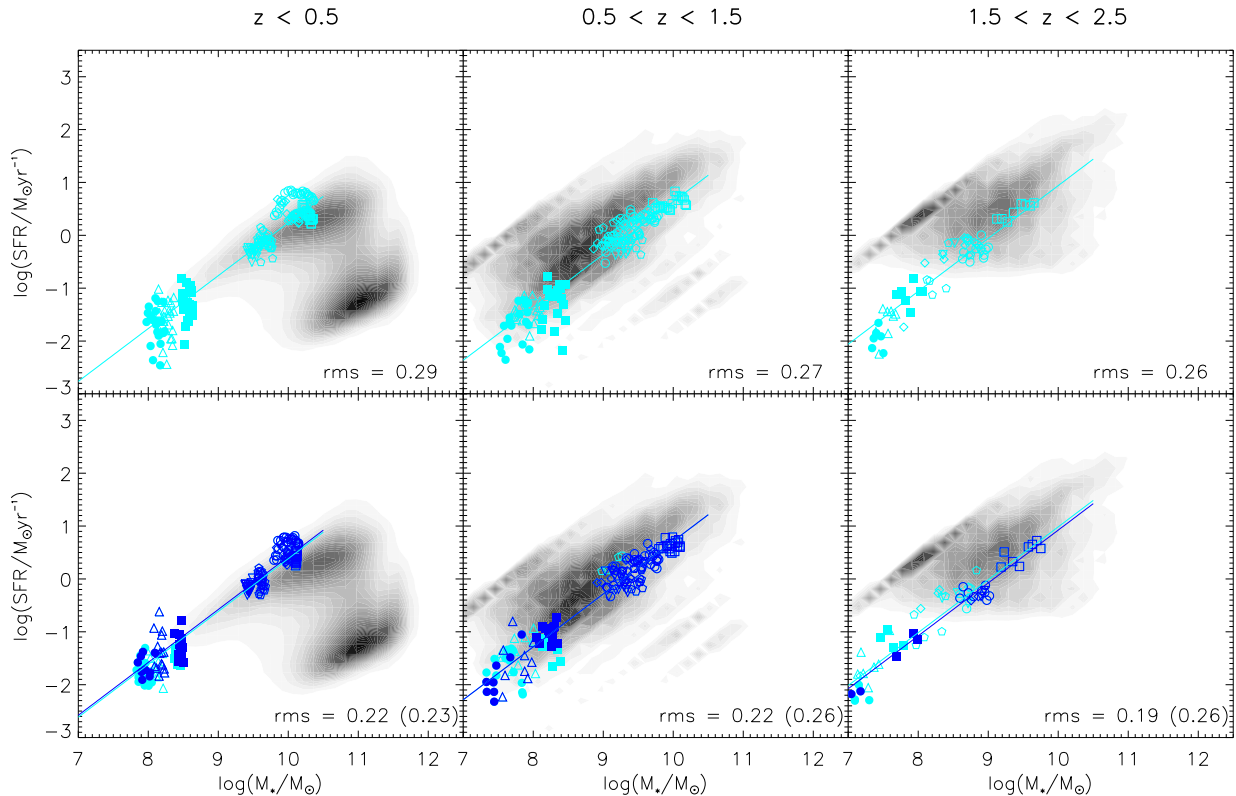}
\caption{
The evolution of M$_{*}$\,-\,SFR. Top panels: SFR and M$_{*}$ values are derived from simulations directly. Bottom panels:  stellar masses were computed using  B-band mass-to-light ratios, 
and the SFR using the TIR-corrected FUV luminosities.
Cyan points correspond to the \textit{complete} sample, while  blue points represent the \textit{limited} one, including only simulations with observational properties within the ranges 
used to define the stellar mass and SFR tracers \citep{Hao:2011,McGaugh:2013}. The shape coding is the same as in Figure~\ref{abundancematching}.
The lines are  linear fits of constant slope 1 which minimize the scatter. 
The dispersions around each line are also shown in each corresponding panel. In the bottom panels, the two values of dispersion correspond to the \textit{limited} and \textit{complete} samples, respectively.
The grey scale contours beneath the simulation data are the normalized 2D histograms of the observational sample of \citealt{Wuyts:2011}. 
In the lowest redshift bin (left panels) the upper overdensity, along which our simulations fall, corresponds to the locus of star forming galaxies having low S\'{e}rsic indices 
\citep[see Figure 1 of][]{Wuyts:2011}.
}
\label{mass_sfr}
\end{figure*} 

\subsection{Evolution of the M$_*$ - SFR relation}
\label{main-sequence-evol}

The correlation between stellar mass and SFR in galaxies across the cosmic time provides a powerful constraint on the star formation implementations used in simulations. 
This correlation reflects the mass build-up of galaxies and their morphological diversity. Observationally, the star forming galaxies have been shown to cluster along the 
so-called \textit{star formation main sequence} \cite[][among others]{Brinchmann:2004,Elbaz:2007,Noeske:2007,Daddi:2007, Karim:2011, Bouwens:2011, Wuyts:2011, Whitaker:2012}.
  This sequence is usually parametrized with a power law ($SFR=\alpha M_{*}^\beta$), where both $\alpha$ and $\beta$
can be taken to vary with the redshift. Typical values of the slope range from $\sim$0.4 to 1 \citep[e.g.][]{Karim:2011,Wuyts:2011,Whitaker:2012,Zahid:2012}. 

Some studies indicate that the slope remains constant,
while  the zero point varies with  redshift. For example, \cite{Zahid:2012} found that a constant slope $\beta=0.7$ with $\alpha$$\sim$$exp(1.33z)$ 
reproduces the observational data up to $z$$\sim$$2.2$. Based on one of the largest mass-complete sample of galaxies up to 
z=2.5, \cite{Whitaker:2012} conclude that the slope of the MS does vary with $z$, but much less than the normalization. 
Most relevant to this present study of simulated  star forming galaxies, they  find a difference when restricting the sample to  the blue star forming galaxies, 
for which they find that the slope is approximately 1. Similarly, \cite{Wuyts:2011} use a large multi wavelength dataset, based on FIREWORKS \citep{Wuyts:2008}, and find that a unitary slope 
reproduces well the observations of (star forming) disc galaxies, where such galaxies are selected as those having low (n$<$2) S\'{e}rsic indices.  
When fitting a slope of 1, they find that the \textit{rms} scatter of the MS is only slightly larger than when both slope and normalizations are left to vary.

Studying the tightness of the MS, \cite{Salmi:2012} also found an intrinsic slope close to unity, 
taking into account the effects of mass measurement errors, larger ranges for the SFR at fixed 
stellar mass and sample completeness. 
Recent studies have proven that large variations in the MS slope are, to a great extent, an effect of sample selection \citep{Salmi:2012,Whitaker:2012,Guo:2013}.

In Figure~\ref{mass_sfr} we show star formation rate as a function of stellar mass for three redshift ranges in the simulated galaxies.
The upper panels give these quantities as derived directly from the simulations (total stellar mass within R$_{lim}$ and SFR computed with the stellar particles 
in the same region that formed in a time range of 100 Myrs). The lower panels show the SFR based on the tracer of \cite{Hao:2011} vs M$_{*}$, calculated using the 
\textit{M/L} ratios in the B-band. 
The blue points represent only the data that fall within the observational ranges of both \cite{Hao:2011} and \cite{McGaugh:2013}, 
while the cyan ones give the whole simulation sample.  
The blue and cyan lines are the linear regressions through the corresponding data points. 
When letting both parameters of the MS vary we obtain slopes in the range $1.0 - 1.1$. 
In the plot we fixed the slope to $1$, and explore the scatter. The MS scatter in this case (the \textit{rms} values shown in each panel) 
is very similar to the one obtained when both slope and normalization are free. In all panels, simulation data has been super-imposed on the data of \cite{Wuyts:2011} (Figure 1 of their paper).

As shown in Figure~\ref{mass_evol}, there are small differences between the mass derived using a tracer based on the SED and the real stellar mass. 
Also the light weighted SFR (blue curves in Figure~\ref{sfrh_evol}) slightly overestimates the \textquoteleft true\textquoteright ~SFR at higher $z$, the effect being larger for the least massive systems.
This translates into a smaller dispersion of SFR at fixed M$_{∗}$ , as can be seen in Figure~\ref{mass_sfr} by comparing the lower with the upper panels. 
As a consequence, the MS scatter is slightly smaller in the \textquoteleft observed\textquoteright ~trends than the \textquoteleft true\textquoteright ~simulation values.

Comparing SFR\,vs\,M$_{*}$ with the observations of \cite{Wuyts:2011} (Figure~1 of their paper), 
we see that at low redshift ($z < 0.50$) simulated galaxies populate a MS very close to the extrapolation 
of the observational sequence. However, at higher redshift the SFRs of simulated galaxies are {\it below} the extrapolations of the observational relations by $\sim~0.4~-~0.5~dex$.
It is important to keep in mind that observational samples are only complete at high stellar masses (e.g. M$_{*}$ $>$ 10$^{9.7}$M$_\odot$).
If we compare the ranges spanned by the simulations with their data set, we see that the simulations fall well 
within the range occupied by observed disc galaxies with S\'{e}rsic indices $n < 2.0$ (bottom panels of Figure~\ref{mass_sfr}).
Our simulations indicate that low mass star forming galaxies should lie below a simple extrapolation of the relation as derived 
from high mass galaxies at high redshift.  
On the other hand, if we compare our data with the MS derived by \cite{Guo:2013} in the redshift range 0.6 - 0.8, the relation in simulations 
for $0.5 < z < 1.5$ is $\sim$ 0.5 dex {\it above} the observational one.

Due to their shallow potential well, low mass galaxies are affected to a greater extent by stellar winds, and have burstier star formation histories.  However, detection 
of such systems is not trivial beyond the local universe. Recently, \cite{Henry:2013a, Henry:2013b} presented a sample of 18 galaxies 
with stellar masses  M$_{*}$$<$$10^{9}$M$_{\odot}$ at intermediate redshift ($z$$\sim$$0.6 - 0.7$), with reliably 
measured emission line fluxes allowing determination of stellar masses and SFRs from SED fitting. 
Most of these galaxies have slightly higher SFR values than our simulations, and a scatter with respect to the linear regression (blue line) in the bottom central panel of Figure~\ref{mass_sfr} of 0.6~dex. 
Recently, \cite{Amorin:2014} presented a sample of star forming dwarfs up to $z\sim0.9$, reaching stellar masses as low as $10^{7}$M$_{\odot}$. As in the \cite{Henry:2013a}, these galaxies have slightly 
higher SFR than our simulations.
It is not yet clear whether the higher observed SFRs of this sample is simply a result of the bias toward observing galaxies with high SFRs.

At even higher redshifts, the only available detection method is galaxy lensing \citep{Richard:2011, Wuyts:2012, Christensen:2012, Belli:2013} allowing the  observation of distant low mass, relatively low SFR galaxies. 
These lensed galaxies at $z > 1.5$ have systematically higher SFR than our simulations, typically an order of magnitude higher, but again it is not clear whether there is a selection bias occurring, given the small number of such observed systems.

\begin{figure}
\centering
\includegraphics[]{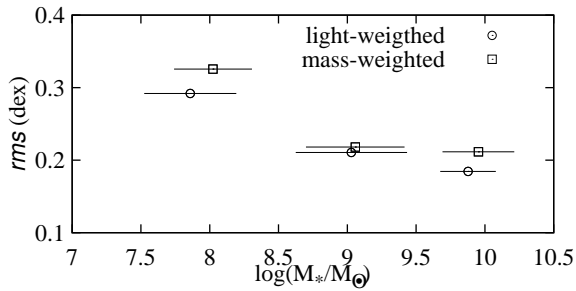}
\caption{
The scatter of the M$_{*}$ - SFR relation for the simulated galaxies. 
The x-error bars represent the stellar mass dispersions. Open squares and circles correspond to the mass-weighted sample (cyan points in the top panels of Figure~\ref{mass_sfr})
and to the light-weighted one (cyan points in the bottom panels of the same figure). 
Since the scatter values have been computed taking into account the evolution of the MS zero point, the data has not been binned in redshift.
For the two samples, a clear trend of decreasing scatter with increasing mass can be appreciated.} 
\label{mass_sfr_scatter}
\end{figure}

\subsection{Scatter in the M$_*$ - SFR relation}
\label{scatter}

The evolution of the star formation main sequence strongly constrains galaxy mass assembly. 
However, it is not only the relation in itself that is important, but also its scatter which reflects the intrinsic variability in the SFR of the star forming galaxy population.  

The star formation rate history is anything but smooth, the variability increasing with decreasing stellar mass
and increasing redshift (see Figure~\ref{sfrh_evol}). The high SFR variability of the lower mass systems is a direct consequence of the shallow potential well of these objects: 
supernovae can  eject large quantities of gas from the galaxies, with star formation cyclically quenched due to the depletion of the cold gas reservoir \citep{Stinson:2007}. 
As the mass of simulated galaxies increases, the triggering and quenching of star formation due to gas outflows  becomes less pronounced.

An advantage of our method of rescaling initial conditions is that we can compare galaxies with different mass that have the same accretion history. Even though they share the same accretion history, 
it can be seen that  g1536-Irr has a significantly more bursty star formation history than g1536\_L$^*$, 
particularly at later times when the potential well of g1536\_L$^*$ is deeper, but also coupled to the lower gas fraction as the more massive galaxy has been more efficient at converting its 
gas into stars, as is evident in Figure~\ref{abundancematching}.

We do not have enough simulated galaxies to derive a statistically robust scatter in the M$_{*}$-SFR relation at a given redshift. 
However, we can test how much scatter comes from the variation in SFR of individual galaxies, as they undergo bursts of star formation and relatively quiescent periods in their evolution. 
In order to derive such scatter, we have to include data from different time steps within the same $z$-bin, 
meaning that we must be mindful that the scatter that we measure is not simply the result of the evolution of the relation itself.
The regular evolution of the relation (the zero point varies linearly with the expansion factor) allows us to find an expected SFR at any $z$ for a given mass, 
and this can be used to derive the scatter in the MS.
In Figure~\ref{mass_sfr_scatter} we show the scatter derived using the expected SFR, as explained above, 
using the mass- and light-weighted stellar masses and star formation rates (open squares and open circles) plotted in the top and bottom panels of Figure~\ref{mass_sfr}, respectively. 
In order to get an estimate of the scatter variation with mass, we divided the two samples into three equally populated stellar mass bins.

We are able to compute the scatter irrespective of redshift
precisely because the reference values for SFR are computed by interpolating the MSs in the three redshift bins.
A clear trend of the MS scatter with stellar mass can be appreciated, less massive galaxies having a higher SFR variability. There is a  difference between the scatter 
computed directly from the simulation outputs (mass-weighted M$_*$ and SFR) and from the \textquoteleft observed \textquoteright fluxes (light-weighted M$_*$ and SFR). This difference
was expected in light of Figure~\ref{sfrh_evol}, where we saw that the SFR tracer better follows the \textit{true} simulation values at high stellar masses than at the low end. 

For the light-weighted sample, the scatter goes from $\sim$0.30 dex for M$_{*}$ $<$ 10$^{8}$M$_\odot$ to $\sim$0.20 dex for M$_{*}$ $>$ 10$^{9}$M$_\odot$
For the more massive simulated galaxies which have M$_{*}$ in the range where observational samples are complete, 
the scatter of $\lesssim 0.3$ is similar to the one found by observers \citep[e.g.][]{Daddi:2007,Noeske:2007,Elbaz:2007,Zahid:2012}. 
Therefore, from our small sample of simulated galaxies, it appears that most and perhaps all of the dispersion in the MS could be explained by the intrinsic scatter in the SFR, 
which is a direct consequence of the bursty star formation history of individual galaxies.

\subsection{The mass-metallicity relation}
\label{massmet}

\begin{figure}
\centering
\includegraphics[]{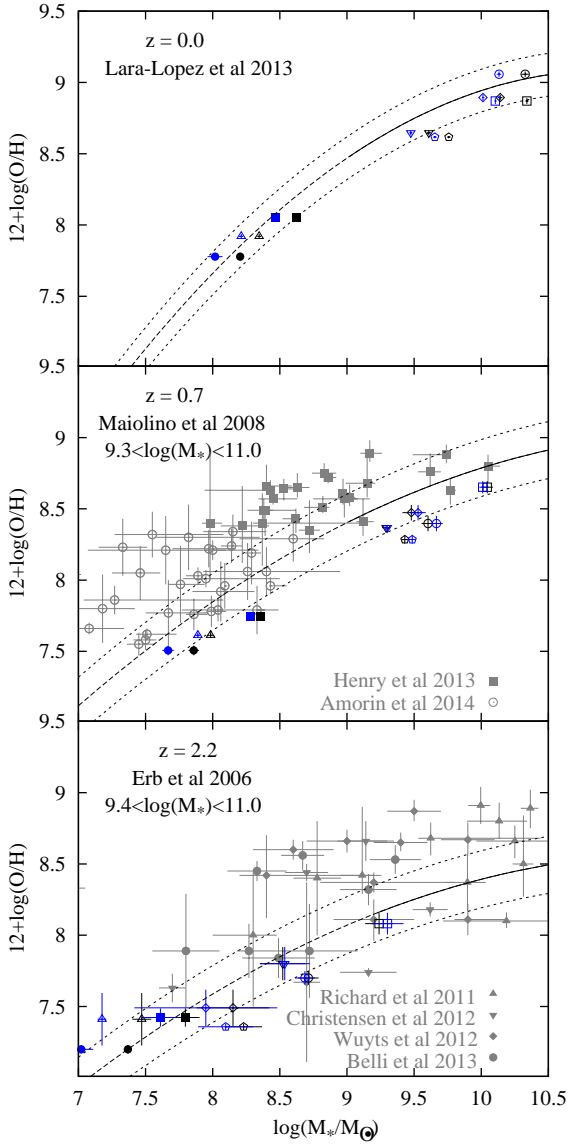}
\caption{Oxygen abundance as function of stellar mass at three different redshifts. 
The blue and black points correspond to masses derived using the mass-to-light ratio in the B band and to \textit{true} simulation values, respectively.
The shape coding is the same as in Figure~\ref{abundancematching}. In the center and bottom panels we over plotted in grey some of the observational samples 
(indicated in the corresponding legends) which have the lowest mass system detected at intermediate and high $z$.
The black solid curves are the observational M-Z relations obtained by \citealt{LaraLopez:2013} (top panel), \citealt{Maiolino:2008} (center panel) and \citealt{Erb:2006} (bottom panel). 
The dashed black lines give the low mass extrapolation of these relations, while the dotted black ones give the scatter of 0.15, 0.20 and 0.20 dex, respectively.
The limits in the center and bottom panels indicate the corresponding stellar mass ranges for the samples of \citealt{Maiolino:2008} and \citealt{Erb:2006}.}
\label{massmetfig}
\end{figure}

Star formation is primarily governed by the cold gas reservoir available. The stellar mass of galaxies correlates with  gas metallicity \citep{Lequeux:1979}.
While the stellar mass reflects the baryon mass locked into stars, the gas metallicity traces the material reprocessed by stars and the coupling of the galaxy to its environment, 
in terms of inflows of fresh or pre-enriched gas and the outflows of metals in galactic winds. 
This correlation provides one of the means by which the feedback processes affecting this baryon cycle can be tested against observations \citep{Brooks:2007}. 

In Figure~\ref{massmetfig} we show the Oxygen abundance versus M$_{*}$ at three redshifts, chosen to make a comparison with the observational samples in \cite{Erb:2006}, 
\cite{Maiolino:2008} and \cite{LaraLopez:2013} at high,  intermediate and low $z$, respectively. 
The M-Z relation of \cite{Maiolino:2008} has been rescaled to a Chabrier IMF.  
We use the mean\footnote{Using the median had minimal impact on the results.} $12+log(O/H)$ of the cold gas ($T < 10^{4}K$) within R$_{lim}$.

The simulated galaxies follow well these observations at high and low redshift (top and bottom panels),
being roughly within the observed scatter from the extrapolations of the M-Z relations \citep{LaraLopez:2013, Erb:2006}.  
For the $z=0$ we use the scatter from \cite{LaraLopez:2013}, while for \cite{Maiolino:2008} and \cite{Erb:2006} curves we assumed a value 0.20~dex. 
At intermediate $z$ there is less agreement in absolute values of simulations and the observations of \cite{Maiolino:2008}, with the former being $\sim~0.2~dex$ 
smaller than the latter in the entire mass range. Yet the trend of metallicity with mass is well reproduced by the simulations even at this intermediate mass range.
In simulations, the metallicity evolution between z$\sim$2.2 and z$\sim$0.7 varies from $\sim$0.25~dex for the lowest mass systems up to $\sim$0.50~dex for the massive ones.
A similar trend can be appreciated in observations, by comparing the extrapolations of the relations in \cite{Erb:2006} and \cite{Maiolino:2008}. 
Thus, the difference in metallicity given by the two curves grows from 0.2~dex for M$_{*}\sim$10$^{7.5}$M$_{\odot}$ to 0.5~dex for M$_{*}\sim$10$^{9.5}$M$_{\odot}$. 
On the other hand, the observational local M-Z relation is steeper than the intermediate and high $z$ ones, 
when considering the extrapolation at lower stellar masses.

In the same figure, we over plotted the data from some  recent observations which reach stellar masses as low as M$_{*}\sim$10$^{7.6}$M$_{\odot}$
at high redshift \citep{Richard:2011, Wuyts:2012, Christensen:2012, Belli:2013}. We also show the samples of \cite{Henry:2013a} at z$\sim$0.65 and of \cite{Amorin:2014} ($z\lesssim 0.9$).
The observational samples at high $z$s have large errors, and generally higher abundances ($\sim$0.2 dex) than the  curve found by \cite{Erb:2006}. 
The three most massive simulated galaxies overlap in the M-Z plane with these observations. The less massive simulated galaxies, on the other hand, 
have smaller metallicities than the \cite{Richard:2011, Wuyts:2012, Christensen:2012, Belli:2013} samples and show up within 2$\sigma$ from the extrapolation of \cite{Erb:2006} relation.
At intermediate redshifts, there is a partial overlap between the \cite{Maiolino:2008} observations and the more recent ones by \cite{Henry:2013a} and \cite{Amorin:2014}, 
although the latter show an offset to higher metallicities with respect to the former. Similar to the high $z$,  the simulations generally have lower metallicities 
than these observations.

It is important to note that different methods are used to measure the metallicities in the different studies, 
meaning that the absolute values are subject to having different calibrations \cite[e.g.][]{Zahid:2012}, 
with different metallicities calibrations resulting in abundances that can vary as much as $\sim$ 0.35 dex.
We also note that the shape of the M-Z relation remains a matter of debate.

\subsection{The mass-SFR-metallicity relation}
\label{msfrmet}

\begin{figure}
\centering
\includegraphics[]{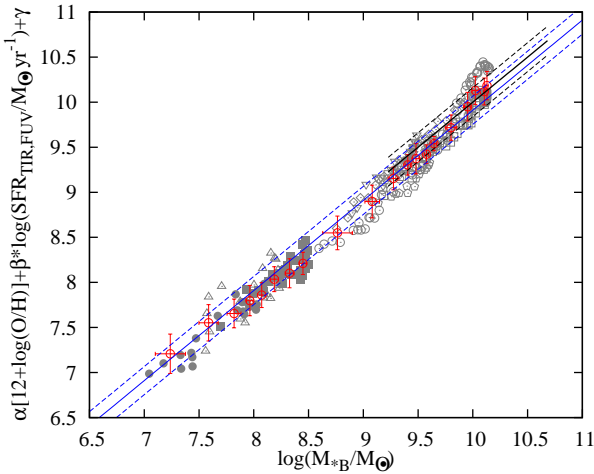}
 \caption{The FMR of star forming galaxies. The stellar mass is given as a linear combination of $12+log(O/H)$ and $log(SFR)$, following \citealt{LaraLopez:2010}, 
  with: $\alpha=1.122$, $\beta=0.474$ and $\gamma=-0.097$. The grey points are the simulated galaxies in the redshift range $z < 3.5$, plotted with the same symbols as in 
  Figure~\ref{abundancematching}. The solid blue line is the linear fit of constant slope 1 which minimize the scatter, while the dashed blue lines give the corresponding 1$\sigma$ 
  of $0.16~dex$. The open red circles represent the simulations in equally populated stellar mass bins. The solid and dashed black lines are the one to one correspondence of
  \citealt{LaraLopez:2010} and its 1~$\sigma$ deviation ($0.16~dex$). The solid blue line is $0.09~dex$ below the solid black one. 
}
\label{mass_sfr_met}
\end{figure}

The correlations found between the stellar mass and SFR \citep{Brinchmann:2004} on one hand, and stellar mass and gas metallicity \citep{Lequeux:1979}, on the other, 
raised the question whether they are not just projections of a fundamental relation. In this respect, \cite{Ellison:2008} found that indeed the mass - metallicity relation 
depends on the SFR. Consequently, in two parallel studies, \cite{Mannucci:2010} and \cite{LaraLopez:2010} proposed combinations of these observables (stellar mass, star formation rate 
and metallicity) which hold across a wide redshift range (up to $z\sim 3.5$).

In order to minimize the scatter of the so called \textit{Fundamental Metallicity Relation} (FMR), \cite{Mannucci:2010} introduced a new variable, $\mu_{\alpha}=log(M_{*})-\alpha\cdotp log(SFR)$.
A value of $\alpha=0.32$ minimizes the scatter of median metallicities of SDSS galaxies, while high $z$ galaxies show the 
same range of $\mu_{0.32}$ as low redshift galaxies. In an equivalent way, \cite{LaraLopez:2010} found that using 
a linear combination of $12+log(O/H)$ and $log(SFR)$, the stellar mass can be recovered with an accuracy of $0.2~dex$.
Therefore, they proposed that $log(M_{*}/M_{\odot})=\alpha[12+log(O/H)]+\beta~log(SFR/M_{\odot}yr^{-1})+\gamma$ with 
a standard deviation of $0.16~dex$, if no errors are considered in metallicity and star formation rate. 
They showed that it holds up to redshift$\sim$3.5.

In Figure~\ref{mass_sfr_met} we give the FMR of our simulated galaxies showing a total of 324 snapshots of our suite of simulations between 
$z=3.5$ and $z=0$ (grey points), and find that they lie  on a linear sequence. 
We fitted a line through the points allowing both normalization and slope to vary obtaining a scatter only $0.01~dex$ smaller
than when fixing the slope to $1$. Thus, we show in solid blue the FMR obtained by setting the slope to 1 and leaving the zero point as only free parameter. 
The dashed blue line is the  corresponding $1\sigma$ scatter of $0.16~dex$. 
The FMR of simulated galaxies is $\sim 0.1~dex$ below the extrapolation of the observational relation to low mass and low metallicity galaxies 
(solid black line, \citealt{LaraLopez:2010}). 
Within the mass range where the simulations overlap with the \cite{LaraLopez:2010} sample, the simulations fall within the 1 $\sigma$ scatter of the observations.     

While both metallicity and SFR correlate tightly with stellar mass, they correlate loosely, with each other, 
in simulations and observations alike. In simulations, stellar mass and cold gas metallicity evolve smoothly and in a quite similar manner with redshift, 
leading to a small scatter in the M-Z relation, while the SFR shows large variations, especially for less massive systems. 
On the other hand, the metallicity appears to correlate with the specific SFR, but with dwarfs evolving differently than massive galaxies.
However, combining the three in the FMR, no redshift evolution is observed and the scatter is reduced with respect to the M$_{*}$-SFR relation.
The emergence of this plane in the space of M$_{*}$-SFR-Z is partially explained by the fact that metallicity increases with decreasing $z$, while 
the SFR shows the opposite behaviour. Therefore, to some degree, the redshift dependences of metallicity and SFR compensate each other. 
We stress, though, that the large dynamical range in this study is obtained by combining snapshots of a small number of galaxies 'observed' at
different redshifts. From this perspective, the simulated galaxies evolve along the FMR. 

\section{Conclusions}
\label{discuss}

Using cosmological galaxy simulations from the MaGICC project, we study the evolution of the stellar masses, star formation rates 
and gas phase abundances of star forming galaxies.  
We derive the stellar masses and star formation rates using observational relations based on spectral energy distributions 
by applying the new radiative transfer code GRASIL-3D to our simulated galaxies. 

We compare light-weighted masses and star formation rates with the their mass-weighted counterparts from simulations. 
We find that the simple stellar mass tracers from \cite{McGaugh:2013}, based on integrated colors and mass-to-light ratios 
are in good agreement with the simulation values, the difference between them being less than $\sim$0.20 dex.
Also, the \cite{Hao:2011} star formation rate tracer based on IR-corrected FUV luminosity follows closely the simulation. 

Although the MaGICC simulations have the stellar feedback fixed in order match  the stellar mass-halo mass relation at a given mass and at $z=0$, 
all our eight galaxies, spanning two 
orders of magnitude in mass, actually are within one $\sigma$ of the abundance matching curves up to redshift $z\sim3.5$. 

We show that the simulated galaxies populate projections of the stellar mass - star formation rate - metallicity
plane, similar to observed star forming disc galaxies. Thus, in the $log(M_{*})$-$log(SFR)$ plane, simulated galaxies fall along a line of slope 1, up to redshift 2.5, 
the normalization increasing with $z$. Both the slope as well as the scatter around these relations are in agreement with observational data. However, the normalizations for the 
simulated galaxies at higher $z$s are lower than in observations by as much as $\sim0.5~dex$. 
This may be a problem for the simulations, yet we emphasize that the simulations are generally of lower mass than complete galaxy surveys, albeit with overlap. 
A close comparison with \cite{Wuyts:2011} data shows that our suite of simulations fall well within the range of observed data. 
Indeed, there is a  hint that the low mass end of observed data has a lower normalization than the high mass end. 
It will be intriguing to see whether this prediction of our model matches future observational samples that are able to probe to lower stellar  masses. 
On our end, a larger sample of simulated galaxies will also allow a more comprehensive comparison of our theoretical model with observations. 
 
Similar to observations, the scatter in the MS seems to be independent of $z$, and decreasing with increasing stellar mass. 
By analyzing this sample of eight galaxies, \textquoteleft observed\textquoteright~ at different times, we conclude that most of the \textit{Main Sequence} scatter can be explained 
by the intrinsically  bursty history of star formation.
 
In the stellar mass - gas phase abundance plane, the simulated discs follow the observations roughly within 2$\sigma$ up to redshift $z\sim2.2$. 
Finally, our sample of simulated star forming galaxies also show a tight correlation among stellar mass, star formation rate and metallicity, with the same coefficients as in 
\cite{LaraLopez:2010} apart from a difference of $0.09~dex$ in normalization. The scatter around this FMR is the same as in the observational sample ($0.16~dex$).
The fact that the FMR does not evolve with redshift for our small sample of simulated galaxies, 'observed' over a wide range of $z$s ($<$3.5), 
shows that our galaxies grow in  away such that the metallicity and SFR evolve in a manner that 'conspires' to maintain the same relation between M$_{*}$-Z-SFR. 
The fact that observed galaxies also evolve in a similarly conspiratorial manner provides support for the baryon cycle within our model. 
The manner in which the M$_{*}$-Z relation depends on SFR at any given redshift will be examined in a future study, 
using a larger sample of simulated galaxies which will allow galaxies of similar mass to be compared at a given redshift.

\section*{Acknowledgements}

We would like to thank Stijn Wuyts for providing the observational FIREWORKS data used
in Figure~\ref{mass_sfr} and to Arjen van der Wel for useful conversations.
This work was partially supported by the MICINN and MINECO (Spain) through the grants
AYA2009-12792-C03-02 and AYA2012-31101 from the PNAyA, as well as by 
the regional Madrid V PRICIT program through the ASTROMADRID network 
(CAM S2009/ESP-1496) and the ''Supercomputaci\'on y e-Ciencia'' 
Consolider-Ingenio CSD2007-0050 project.
We also gratefully acknowledge the computer resources provided by BSC/RES (Spain), the Centro de Computaci\'on Cientif\'ica (UAM, Spain), 
STFCs DiRAC Facility (through the COSMOS: Galactic Archaeology programme), 
the DEISA consortium, co-funded through EU FP6 project RI-031513 and the FP7 project RI-222919 
(through the DEISA Extreme Computing Initiative), the PRACE-2IP Project (FP7 RI-283493), 
and the University of Central Lancashire’s High Performance Computing Facilty.
A. Obreja is supported by MICINN and MINECO (Spain) through a FPI fellowship.
C. Brook thanks MINECO for financial support through contract associated to AYA2009-12792-C03-03 grant. 
G. Stinson received funding from the European Research Council under the European Union's 
Seventh Framework Programme (FP 7) ERC Grant Agreement n. 321035.

\bibliographystyle{mn2e}
\bibliography{aobreja_v2}

\label{lastpage}

\end{document}